\documentclass[12pt,tightenlines]{revtex4}
\usepackage{amsfonts}
\usepackage{amsmath}
\usepackage{txfonts}
\usepackage{graphicx}
\usepackage{amssymb}
\usepackage{color}
\usepackage{hyperref}
\usepackage{wrapfig,lipsum,booktabs}
\usepackage{floatrow,float}
\usepackage[english]{babel}

\begin{document}
\title{Discrete structure of the brain rhythms}
\author{L. Perotti$^{1}$, J. DeVito$^{2}$, D. Bessis$^{1}$, Y. Dabaghian$^{2*}$}
\affiliation{$^1$Department of Physics, Texas Southern University, 3100 Cleburne Ave., Houston, Texas 77004, \\
$^2$Department of Neurology, The University of Texas Health Science Center at Houston, Houston, TX 77030\\
$^{*}$e-mail: yuri.a.dabaghian@uth.tmc.edu}
\vspace{17 mm}
\date{\today}
\vspace{17 mm}
\begin{abstract}
Neuronal activity in the brain generates synchronous oscillations of the Local Field Potential 
(LFP). The traditional analyses of the LFPs are based on decomposing the signal into simpler components, such 
as sinusoidal harmonics. However, a common drawback of such methods is that the decomposition primitives are 
usually presumed from the onset, which may bias our understanding of the signal's structure. Here, we introduce 
an alternative approach that allows an impartial, high resolution, hands-off decomposition of the brain waves into 
a small number of discrete, frequency-modulated oscillatory processes, which we call oscillons. In particular, we 
demonstrate that mouse hippocampal LFP contain a single oscillon that occupies the $\theta$-frequency band and 
a couple of $\gamma$-oscillons that correspond, respectively, to slow and fast $\gamma$-waves. Since the oscillons 
were identified empirically, they may represent the actual, physical structure of synchronous oscillations in neuronal 
ensembles, whereas Fourier-defined ``brain waves'' are nothing but poorly resolved oscillons.
\end{abstract}
\maketitle
%%%%%%%%%%%%%%%%%%%%%%%%%%%%%%%%%%%%%%%%%%%%%%%%%%%%%%%%%

%\section*{Author Summary} We propose a novel approach to analyzing biological signals, in which the signal's 
%prime components are not a priori presumed, but discovered empirically. By applying this method to the analysis 
%of the hippocampal ``brain waves,'' we demonstrate that the latter decompose into a superposition of a few phase 
%modulated oscillatory processes, which we call \emph{oscillons}. We hypothesize that oscillons represent the actual, 
%physical structure of the LFP rhythms, whereas Fourier-defined brain waves are nothing but poorly resolved oscillons. 
%Identification of these structures may help connecting experimental data with theoretical models of the neuronal 
%synchronization.
%For example, the $\theta$-oscillon corresponds to the standard, Fourier-defined $\theta$-rhythm and a 
%couple of $\gamma$-oscillons correspond to slow and fast $\gamma$-rhythms. 

%Running title: Oscillons in the brain rhythms
\section{Introduction}
\label{section:intro}

Neurons in the brain are submerged into a rhythmically oscillating electrical field, created by synchronized 
synaptic currents \cite{Buzsaki1}. The corresponding potential, known as local field potential (LFP) is one 
of the principal determinants of neural activity at all levels, from the synchronized spiking of the individual 
neurons to high-level cognitive processes  \cite{Thut}. The attempts to understand the structure and function 
of LFP oscillations, and of their spatiotemporally smoothed counterparts---the electroencephalograms (EEG), 
continues for almost a century and a systematic understanding of their roles begins to shape. 

The possibility to identify true physiological functions of the LFP depends fundamentally on the mathematical 
and computational tools used for its analysis. The majority of the currently existing methods are based on 
breaking the signal into a combination of simpler components, such as sinusoidal harmonics or wavelets 
\cite{Boashash,Vugt}, and then correlating them with physiological, behavioral and cognitive phenomena 
\cite{Kopell,BuzsakiBook}. For example, wavelet analysis is most appropriate for studying time-localized 
events, such as ripples or spindles \cite{Battaglia,Sitnikova}, whereas for the general analyses, the oscillatory 
nature of LFPs suggests using Fourier decomposition into a set of plane waves with a fixed set of frequencies 
$\omega, 2\omega, 3\omega, \ldots$. The latter approach has dominated the field for the last several decades 
and now constitutes, in effect, the only systematic framework for our understanding of the structure and the 
physiological functions of the brain rhythms \cite{BuzsakiBook}.
However, a common flaw of these methods is that the decomposition primitives are presumed from the onset, 
and the goal of subsequent analyses reduces merely to identifying the combination that best reproduces the 
original signal. Since no method can guarantee a universally good representation of the signals' features and 
since the physiological structure of the LFPs remains unknown, obtaining a physically adequate description of 
the brain rhythms is a matter of fundamental importance.

Below we propose a novel approach of LFP analysis based on a recent series of publications \cite{Bessis1,Bessis2,Perotti1}, 
in which an optimal set of frequencies  $\omega_1, \omega_2, \ldots$, is estimated, at every moment of time 
$t$, using the Pad\'{e} Approximation Theory \cite{Baker}. In contrast with the Fourier method, these adaptively 
optimized values can freely change within the sampling frequency domain, guided only by the signal's structure. 
The resulting harmonics are highly responsive to the signals' dynamics and capture subtle details of the signal's 
spectrum very effectively, as one would expect from a Pad\'{e} Approximation based technique. We call the new 
method Discrete Pad\'{e} Transform (DPT), to emphasize certain key correspondences with the traditional Discrete 
Fourier Transform (DFT).

Applying DPT analyses to LFP rhythms recorded in mouse hippocampi reveals a new level in their structure--a 
small number of frequency-modulated oscillatory processes, which we call \emph{oscillons}. Importantly, oscillons 
are observed in the physiologically important theta ($\theta$) \cite{BuzsakiTh1,BuzsakiTh2,Arai} and gamma 
($\gamma$) \cite{ColginGamma,Basso} frequency domains, but are much sharper defined. For example, in the 
Fourier approach, the $\theta$-rhythm is loosely defined as a combination of the plane waves with frequencies 
between $4$ and $12$ Hz \cite{BuzsakiTh1,BuzsakiTh2,Arai}. In contrast, our method suggests that there exists 
a \emph{single} frequency-modulated wave---the $\theta$-oscillon---that occupies the entire $\theta$ frequency 
band and \emph{constitutes} the $\theta$-rhythm. Similarly, we observe oscillons in the low and high 
$\gamma$-frequency domains. The superposition of the oscillons reproduces the original LFP signal with high 
accuracy, which implies that these waves provide a remarkably sparse representation of the LFP oscillations. 
Since oscillons emerged as a result of empirical analyses, we hypothesize that they represent the actual, physical 
structure of synchronized neuronal oscillations, which were previously approximately described as the Fourier-defined 
``brain waves.''

\section{Results}
\label{section:results}

\textbf{The oscillons}. We implemented a ``Short Time Pad\'{e} Transform'' (STPT), in which a short segment of 
the time series (that fits into a window of a width $T_W$) is analyzed at a time. This allows us to follow the signal's 
spectral composition on moment-to-moment basis and to illustrate its spectral dynamics using Pad\'{e} spectrograms 
(the analogues of to the standard Fourier spectrograms \cite{Jacobsen,Howell}). 

\begin{figure}
\includegraphics[scale=0.72]{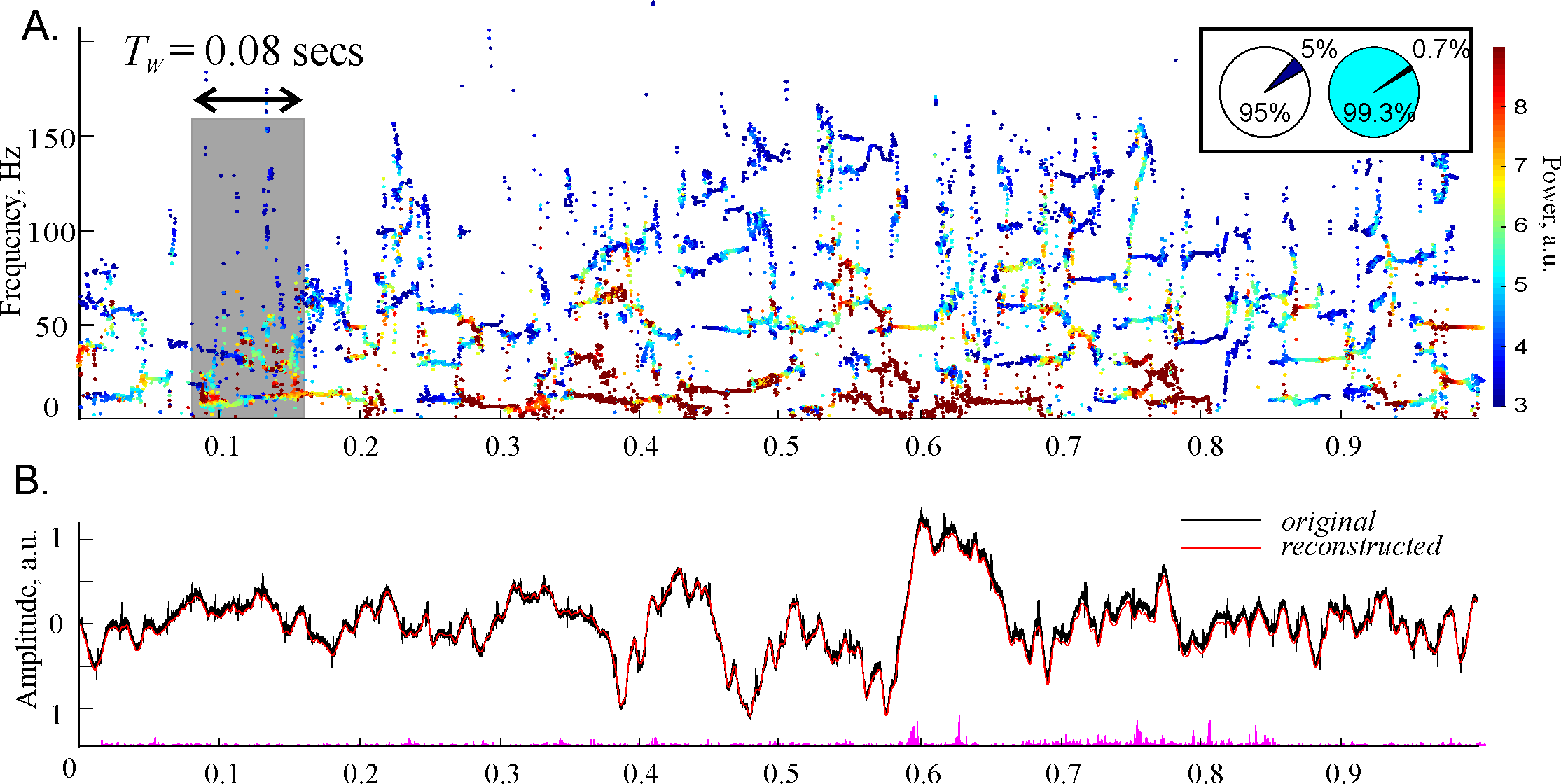}
\caption{{\footnotesize \textbf{Pad\'{e} spectrograms of the hippocampal LFP signal}. \textbf{A}. Discrete Pad\'{e} 
Spectrogram (DPS) produced for the LFP signal recorded in the CA1 region of the rodent hippocampus at the sampling 
rate $10$ kHz. At each moment of time, the vertical cross section of the spectrogram gives the instantaneous set of 
the regular frequencies. At consecutive sequence of moments of time, these frequencies produce distinct, contiguous 
traces, which can be regarded as timelines of discrete oscillatory processes---the spectral waves with varying 
frequencies $\omega_q(t)$, amplitudes $A_q(t)$ (shown by the color of dots) and phases $\psi_q(t)$ (not shown). 
Note that the higher frequency spectral waves tend to have lower amplitudes. Highest amplitudes appear in the 
$\theta$-region, i.e. in the frequency range between $4$ and $12$ Hz. The spectral waves above $100$ Hz tend to 
be scarce and discontinuous, representing time-localized splashes of LFP. The width of the time window is $T_W  = 
0.08$ sec ($800$ data points). The pie diagrams in the box show that stable harmonics constitute only 5\% of their 
total number, but carry over 99\% of the signal's power. \textbf{B}. The LFP signal reconstructed from the regular 
poles (red trace) closely matches the original signal (black trace) over its entire length, which demonstrates that the 
oscillon decomposition (\ref{oscdec}) provides an accurate representation of the signal. The difference between the 
original and the reconstructed signal is due to the removed noise component---the discarded ``irregular'' harmonics 
(the magenta ``grass'' along the $x$-axis). Although their number is large (about $90-99\%$ of the total number of 
frequencies), their combined contribution is small---only about $10^{-3}-10^{-4} \%$ of the signals power.}}
\label{Figure1}
\end{figure}

Applying these analyses to the hippocampal LFPs recorded in awake rodents during habituation stage \cite{Tang}, we 
observed that there exist two types of time-modulated frequencies (Fig.~\ref{Figure1}). First, there is a set of 
frequencies that change across time in a regular manner, leaving distinct, continuous traces---the \emph{spectral waves}. 
As shown on Fig.~\ref{Figure1}A, the most robust, continuous spectral waves with high amplitudes (typically three or 
four of them) are confined to the low frequency 
domain and roughly correspond to the traditional $\theta$- and $\gamma$-waves \cite{BuzsakiTh1,ColginGamma}. 
The higher frequency (over $100$ Hz) spectral waves are scarce and short, representing time-localized oscillatory 
phenomena that correspond, in the standard Fourier approach, to fast $\gamma$ events \cite{Sullivan}, sharp wave 
ripples (SWRs) \cite{SWR} or spindles \cite{Latchoumane}. 
Second, there exists a large set of ``irregular'' frequencies that assume sporadic values from one moment to another, 
without producing contiguous patterns and that correspond to instantaneous waves with very low amplitudes. 

%%%%%%%%%%%%%%%%%%%%%%%%%%%%%%%%%%%
%%%%%%%%%%%%%%%%%%%%%%%%%%%%%%%%%%%

From the mathematical perspective, the existence of these two types of instantaneous frequencies can be explained 
based on several subtle theorems of Complex Analysis, which point out that the ``irregular'' harmonics represent the 
signal's noise component, whereas the ``regular,'' stable harmonics define its oscillatory part (see \cite{Steinhaus,
Froissart,Gilewicz1,Gilewicz2} and the Mathematical Supplement). Thus, in addition to revealing subtle dynamics the 
frequency spectrum, the DPT method allows a context-free, impartial identification of noise, which makes it particularly 
important for the biological applications \cite{Faisal,Ermentrout}.

As it turns out, the unstable, or ``noisy,'' frequencies typically constitute over 95\% of the total number of harmonics 
(Fig.~\ref{Figure1}A). However, the superposition of the harmonics that correspond to the remaining, \emph{stable} 
frequencies captures the shape of the signal remarkably well (Fig.~\ref{Figure1}B). In other words, although only a 
small portion of frequencies are regular, they contribute over 99\% of the signal's amplitude: typically, the original LFP 
signal differs from the superposition of the stable harmonics by less than $1\%$. If the contribution of the ``irregular'' 
harmonics (i.e., the noise component $\xi(t)$) is included, the difference is less than $10^{-4}-10^{-6}$ of the signal's 
amplitude.

These results suggest that the familiar Fourier decomposition of the LFP signals into a superposition of plane waves 
with \emph{constant} frequencies,
\begin{equation}
r(t) = \Sigma^N_{p =1} a_p e^{i\omega_pt},
\label{fourdec}
\end{equation}
should be replaced by a combination of a few phase-modulated waves embedded into a weak noise background $\xi(t)$,
\begin{equation}
s(t) = \Sigma^M_{q=1} A_q e^{i\phi_q(t)}+\xi(t),
\label{oscdec}
\end{equation}
which we call \emph{oscillons}. 
We emphasize that the number $M \ll N$ of the oscillons in the decomposition (\ref{oscdec}), their amplitudes $A_q$, 
their phases $\phi_q$ and the time-dependent frequencies $\omega_q(t)=\partial_t\phi_q(t)$ (i.e., the spectral waves 
shown on Fig.~\ref{Figure1}A) are reconstructed on moment-by-moment basis from the local segments of the LFP signal 
in a hands-off manner: we do not presume \emph{a priori} how many frequencies will be qualified as ``stable,'' when 
these stable frequencies will appear or disappear, or how their values will evolve in time, or what the corresponding 
amplitudes will be. Thus, the structure of the decomposition (\ref{oscdec}) is obtained \emph{empirically}, which suggests 
that the oscillons may reflect the actual, physical structure of the LFP rhythms. 

\begin{figure}
\includegraphics[scale=0.72]{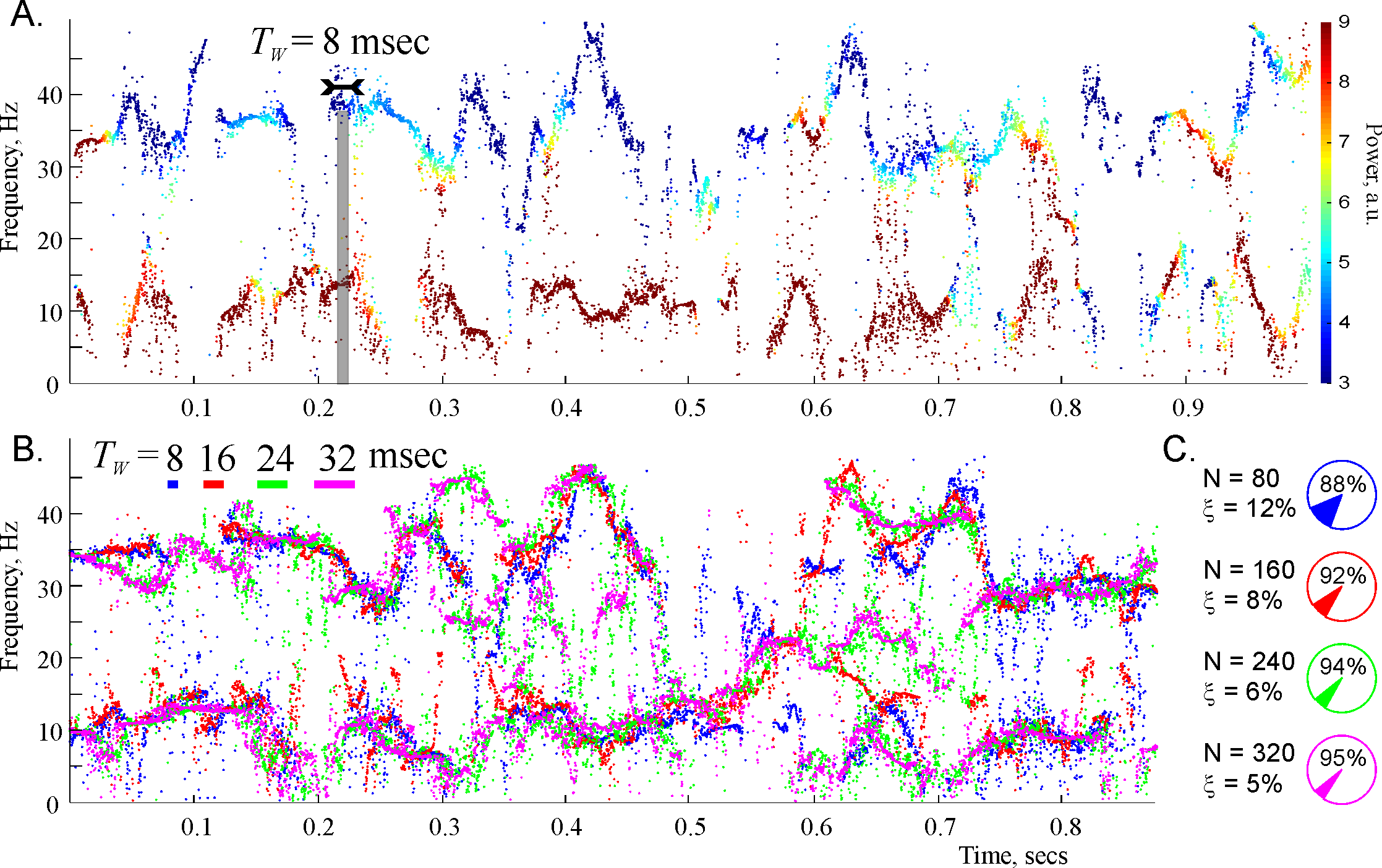}
\caption{{\footnotesize \textbf{Pad\'{e} spectrograms of the rat hippocampal LFP signal}. \textbf{A}. A detailed 
representation of the lower portion the spectrogram recomputed for $T_W  = 0.08$ sec ($80$ data points) 
exhibits clear oscillatory patterns. \textbf{B}. The shape of the two lowest frequency spectral waves is stable 
with respect to the variation of time window size, $T_W$. The strikes of different color in the top left corner 
represent the widths of the four $T_W$-values used in DPT analysis. The corresponding reconstructed frequencies 
are shown by the dots of the same color. Although the frequencies obtained for different $T_W$s do not match 
each other exactly, they outline approximately the same shape, which, we hypothesize, reflects the physical 
pattern of synchronized neuronal activity that produced the analyzed LFP signal. 
\textbf{C}. Pie diagrams illustrate the numbers of data points $N = 80$, $N = 160$, $N = 240$, $N = 320$ 
and the mean numbers of the regular and the irregular (noisy) harmonics in each case.}}
\label{Figure2}
\end{figure}

\textbf{The spectral waves}. We studied the structure the two lowest spectral waves using high temporal 
resolution spectrograms (Fig.~\ref{Figure2}A). Notice that these spectral waves have a clear oscillatory structure,
\begin{eqnarray}
\omega_{q} (t) = \omega_{q,0} + \omega_{q,1} \sin(\Omega_{q,1}t + \varphi_{q,1}) + 
\omega_{q,2}\sin(\Omega_{q,2} t +\varphi_{q,2}) + \ldots, \ \ \ q =1,2,
\label{omt}
\end{eqnarray}
characterized by a mean frequency $\omega_{q,0}$, as well as by the amplitudes, $\omega_{q,i}$, the 
frequencies, $\Omega_{\theta,i}$, and the phases, $\varphi_{\theta,i}$, of the modulating harmonics. The 
lowest wave has the mean frequency of about $8$ Hz and lies in the domain $2 \leq \omega/2\pi \leq 17$ Hz, 
which corresponds to the $\theta$-frequency range \cite{BuzsakiTh1}. The second wave has the mean frequency 
of about $35$ Hz and lies in the low-$\gamma$ domain $25 \leq \omega /2\pi \leq 45$ Hz \cite{ColginGamma}. 
Importantly, the spectral waves are well separated from one another: the difference between the their mean 
frequencies is larger than their amplitudes, which allows indexing them using the standard brain wave notations, 
as $\omega_{\theta}(t)$ and $\omega_{\gamma_{l}}(t)$ respectively, e.g., 
\begin{equation}
\omega_{\theta} (t) = \omega_{\theta,0} + \omega_{\theta,1} \sin(\Omega_{\theta,1}t + \varphi_{\theta,1}) + 
\omega_{\theta,2}\sin(\Omega_{\theta,2} t +\varphi_{\theta,2}) + \ldots \ ,
\label{omgth}
\end{equation}
for the $\theta$ spectral wave an 
\begin{equation}
\omega_{\gamma_l} (t) = \omega_{\gamma_l,0} + \omega_{\gamma_l,1} \sin(\Omega_{\gamma_l,1}t 
+ \varphi_{\gamma_l,1}) + \omega_{\gamma_l,2}\sin(\Omega_{\gamma_l,2} t +\varphi_{\gamma_l,2}) + \ldots 
\label{omgq}
\end{equation}
for the low-$\gamma$ spectral wave, etc.

We verified that these structures are stable with respect to the variations of the SDPT parameters, e.g., to changing 
the sliding window size, $T_W$. The size of the sliding window, and hence the number of points $N$ that fall within 
this window can be changed by over $400\%$, without affecting the overall shape of the spectral waves (Fig.~\ref{Figure2}B). 
The smallest window size (a few milliseconds) is restricted by the requirement that the number of data points captured 
within $T_W$ should be bigger than the physical number of the spectral waves. On the other hand, as the maximal 
value of $T_W$ is limited by the temporal resolution of STPT: if the size of the window becomes comparable to the 
characteristic period of a physical spectral wave, then the reconstructed wave looses its undulating shape and may 
instead produce a set of sidebands surrounding the mean frequency \cite{Boashash}. This effect limits the magnitude 
of the $T_W$ to abut $50$ milliseconds---for larger values of $T_W$, the undulating structure begins to straighten out, 
as shown on  Fig.~\ref{Figure1}A for $T_W=80$ msec.

%%%%%%%%%%%%%%%%%%%%%%%%%%%%%%%%%%%
%%%%%%%%%%%%%%%%%%%%%%%%%%%%%%%%%%%

In contrast with this behavior, the values of the irregular frequencies are highly sensitive to the sliding window size 
and other DPT parameters, as one would expect from a noise-representing component. The corresponding ``noisy'' 
harmonics can therefore be easily detected and removed using simple numerical procedures (see Mathematical Supplement).  
Moreover, we verified that the structure of the Pad\'{e} Spectrogram, i.e., of the parameters the oscillons remain stable 
even if the amount of numerically injected noise exceeds the signal's natural noise level by an order of magnitude (about 
$10^{-4}$ of the signal's mean amplitude), which indicates that the oscillatory part of the signal is robustly identified.

\begin{figure}[ht]
\includegraphics[scale=0.7]{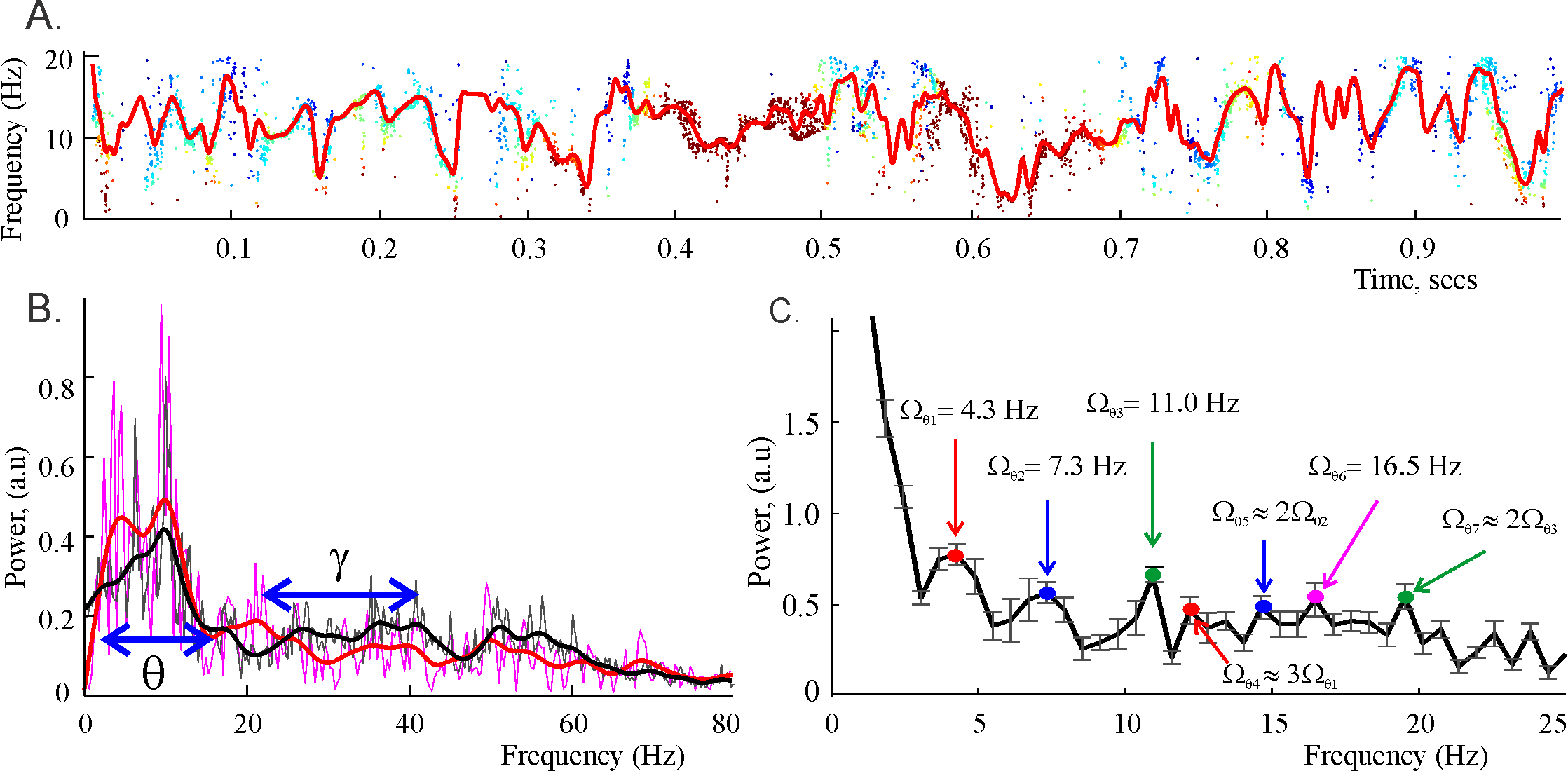}
\caption{{\footnotesize \textbf{Parameters of the spectral waves}. \textbf{A}. The red curve shows the smoothened 
$\theta$ spectral wave, obtained by interpolating the ``raw'' trace of the reconstructed frequencies shown on 
Fig.~\ref{Figure2}A over the uniformly spaced time points. 
\textbf{B}. The power spectra produced by the Discrete Pad\'{e} decomposition (DPT, red) and the standard Discrete 
Fourier decomposition (DFT, black) exhibit characteristic peaks around the mean frequency of the $\theta$-oscillon, 
$\omega_{\theta,0}/2\pi\approx 7.5$ Hz. The height of the peaks defines the amplitudes, respectively, of the 
$\theta$-oscillon in the DPT approach and of the $\theta$-rhythm in DFT. A smaller peak at about $34$ Hz corresponds 
to the mean frequency of the low $\gamma$ oscillon, $\omega_{\gamma_{l,0}}/2\pi \approx 34$. The $\theta$ and 
the low $\gamma$ frequency domains, marked by blue arrows, are defined by the amplitudes of the corresponding 
spectral waves. 
\textbf{C}. The smoothened waves are used to compute the DFT transform and to extract the modulating frequencies 
$\Omega_{\theta,1}\approx 4.3$ Hz, $\Omega_{\theta,2}\approx 7.3$ Hz, $\Omega_{\theta,3}\approx11$ Hz, ..., of 
the decomposition (\ref{omgth}--\ref{omgq}). The error margin in most estimates is $\pm 0.5$ Hz. Notice that there 
exist several approximate resonant relationships, e.g., $\Omega_{\theta,4} \approx 3\Omega_{\theta,1}$, 
$\Omega_{\theta,5}\approx 2\Omega_{\theta,2}$ and $\Omega_{\theta,7}\approx \Omega_{\theta,3}$, which suggest 
that the spectral $\theta$-wave contains higher harmonics of a smaller set of prime frequencies.}}
\label{Figure3}
\end{figure}

\textbf{Parameters of the low frequency oscillons}. To obtain a more stable description of the underlying patterns, we 
interpolated the spectral waves over the uniformly spaced time points (Fig.~\ref{Figure3}A) and then studied the 
resulting ``smoothened'' spectral waves using the standard DFT tools. In particular, we found that, for studied LFP 
signals, the mean frequency of the $\theta$-oscillon is about $\omega_{\theta,0}/2\pi = 7.5 \pm 0.5$ Hz and the 
mean frequency of the low $\gamma$-oscillon is $\omega_{\gamma_{l,0}}/2\pi = 34 \pm 2$ Hz, which correspond 
to the traditional (Fourier defined) average frequencies of the $\theta$ and the low $\gamma$ rhythms. 

%%%%%%%%%%%%%%%%%%%%%%%%%%%%%%%%%%
%%%%%%%%%%%%%%%%%%%%%%%%%%%%%%%%%%

The amplitudes of the $\theta$ and the low $\gamma$ spectral waves---$7.0\pm 1.5$ Hz and $10.1\pm 1.7$ Hz 
respectively---define the frequency domains (spectral widths) of the $\theta$ and the low $\gamma$ rhythms 
(Fig.~\ref{Figure3}B). The amplitudes of the corresponding oscillons constitute approximately $A_{\theta}/A 
\approx 62\%$ and $A_{\gamma_{l}}/A\approx 17\%$ of the net signals' amplitude $A$, i.e., the $\theta$ and the 
low $\gamma$ oscillons carry about $80\%$ of the signals' magnitude.

The oscillatory parts of the spectral waves are also characterized by a stable set of frequencies and amplitudes: for 
the first two modulating harmonics we found $\omega_{\theta,1}/2\pi \approx 4.3$ Hz, $\omega_{\theta,2}/2\pi 
\approx3.2$ Hz for the $\theta$ spectral wave (\ref{omgth}) and $\omega_{\gamma_{l,1}}/2\pi\approx 6.1$ Hz, 
$\omega_{\gamma_{l,2}}/2\pi\approx 4.3$ Hz for the $\gamma$ spectral wave (\ref{omgq}). The corresponding 
modulating frequencies for the $\theta$-oscillon are $\Omega_{\theta,1} = 4.3\pm 0.45$ Hz, $\Omega_{\theta,2} 
= 7.3 \pm 0.48$ Hz, $\ldots$, (Fig.~\ref{Figure3}C). The lowest modulating frequencies for the $\gamma$-oscillon 
are slightly higher: $\Omega_{\gamma_{l},1}= 5.3 \pm 0.41$ Hz, $\Omega_{\gamma_{l},2} = 8.3\pm 0.51$ Hz, 
$\ldots$. In general, the modulating frequencies tend to increase with the mean frequency. 

Importantly, the reconstructed frequencies sometimes exhibit approximate resonance relationships (Fig.~\ref{Figure3}C), 
implying that some of the higher order frequencies may be overtones of a smaller set of prime frequencies that 
define the dynamics of neuronal synchronization \cite{Strogatz,Arenas,Hoppensteadt}.
%($\Omega_{\theta,n> 3}$, $\Omega_{\gamma,n > 4}$)

\section{Discussion}
\label{section:discussion}

\begin{figure}[ht]
\includegraphics[scale=0.72]{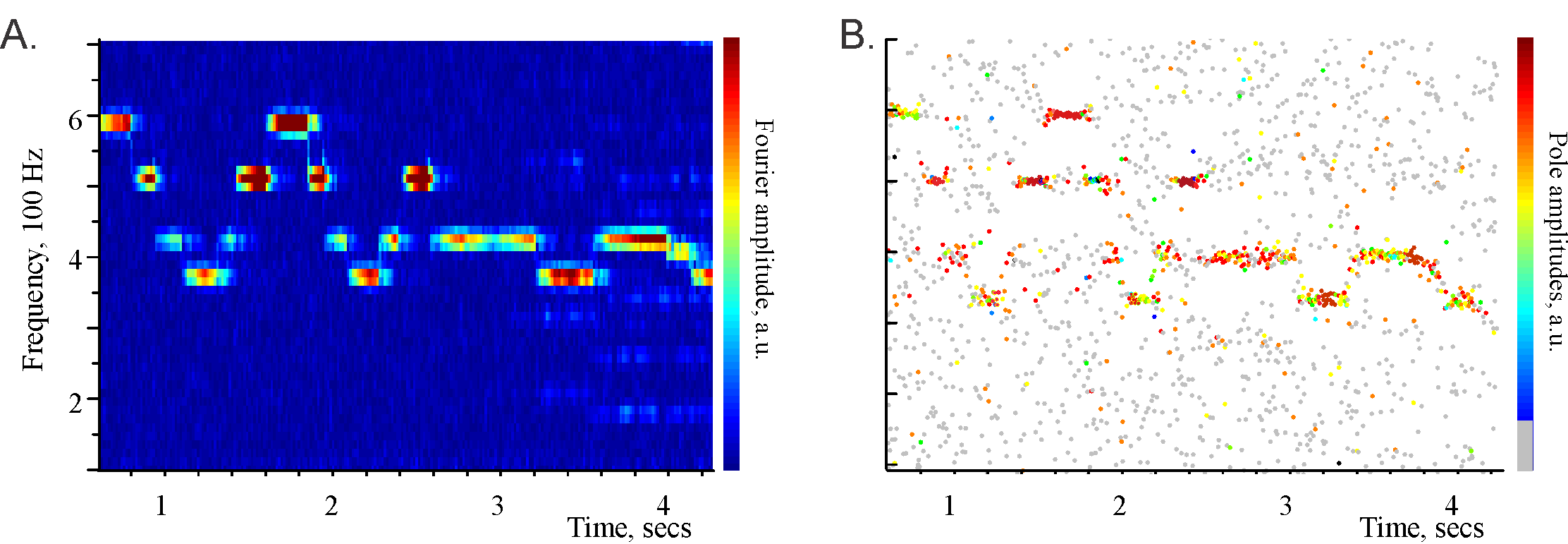}
\caption{{\footnotesize \textbf{Correspondence between the Discrete Fourier (left) and Pad\'{e} (right) spectral 
decompositions}. \textbf{A}. Fourier spectrogram of a $10$ second long excerpt from C. Debussy's Preludes, Book 1: 
No. 8. \emph{La fille aux cheveux de lin}, in which the individual notes are clearly audible. The high amplitude streaks 
(colorbar on the right) correspond to the notes (D\#5, B4, G4, F4, G4, B4, D5, B4, G4, F4, G4, B4, G4, F4, G4, F4, \,...). 
\textbf{B}. The Discrete Pad\'{e} spectrogram of the same signal. The frequencies produced by large amplitude poles 
(see colorbar on the right) match the frequencies of their Fourier counterparts shown on the left. The frequencies 
produced the Froissart doubles form a very low amplitude background ``dust,'' shown in gray. Our main hypothesis 
is that the oscillons detected in the LFP signals by the DPT method may be viewed as ``notes'' within the neuronal 
oscillations.}}
\label{Figure4}
\end{figure}

The Fourier and the Pad\'{e} decompositions agree in simple cases, e.g., both spectrograms resolve the individual 
piano notes in a $10$ sec excerpt from one of Claude Debussy's Preludes (Fig.~\ref{Figure4}).

%%%%%%%%%%%%%%%%%%%%%%%%%%%%%%%%%%%
%%%%%%%%%%%%%%%%%%%%%%%%%%%%%%%%%%%

However, in more complex cases the DTP approach produces a more accurate description of the signal's structure. 
For example, the Pad\'{e} decomposition previously used to detect faint gravitational waves in resonant interferometers, 
which were completely missed by the Fourier analyses \cite{Perotti2}. In the case of the LFP signals, this method 
identifies a small number of structurally stable, frequency-modulated oscillons which may reflect the physical 
synchronization patterns in the hippocampal network. 

Why these structures were not previously observed  via Fourier method? The reason lies in the insufficient resolution 
of the latter, which is due to the well-known inherent conflict between the frequency and the temporal resolutions 
in Fourier Analysis \cite{Grunbaum}. Indeed, in order to observe changes in the signal' spectrum, the size of the 
sliding time window, $T_W$, should be \emph{smaller} than the characteristic timescale of frequency's change, 
$T_W<\Delta T$. On the other hand, reducing $T_W$ implies lowering the number of data points in the sliding 
window, which results in an equal reduction of the number of the discrete harmonics, in both the DFT and the DPT 
approaches. However, since in DFT method these harmonics are restricted to a rigid, uniformly distributed set of 
values (Fig.~\ref{SFigure1}A), a decrease in the number of data points necessarily results in an increase of the interval 
between neighboring discrete frequencies, i.e., in an unavoidable reduction of frequency resolution. In contrast, the 
DPT harmonics  can move freely in the available frequency domain, responding to the spectral structure of the signal 
and providing a high resolution of the signals' spectrum \cite{Perotti1}. In other words, an increase in temporal resolution 
in DPT does not necessarily compromise the frequency resolution and vice versa, which allows describing the signal 
dynamics much more capably. 

In the specific case illustrated on Fig.~\ref{Figure2}, the characteristic amplitudes of the spectral waves is about 
$15-25$ Hz. Producing such frequency resolution in DFT at the sampling rate $S =10$ kHz would require some 
$N = 300-500$ constant frequency harmonics, i.e., $N = 300-500$ data points, which can be collected over $T_W 
= 30-50$ msec time window. However, the characteristic period of the spectral waves is about $60$ msec, which 
implies that for such $T_W$s, the DFT will not be able to resolve the frequency wave dynamics and will replace it 
by an average frequency with some sidebands (see Mathematical Supplement). In contrast, a DPT that uses as few 
as $80$ data points in a $T_W = 8$ msec wide time window, reliably capturing the shape of the spectral wave, which 
then remains overall unchanged as $T_W$ increases fourfold.  
 
Another key property of the DPT method is the intrinsic marker of noise, which is particularly important in biological 
applications \cite{Faisal,Ermentrout}. In general, the task of distinguishing ``genuine noise'' from a ``regular, but highly 
complex'' signal poses not only a computational, but also a profound conceptual challenge \cite{Barone,Shadlen}. In 
contrast with the standard \emph{ad hoc} approaches, the DPT method allows a context-free, impartial identification of 
the noise component, as the part of the signal represented by the irregular harmonics. 

The new structure also dovetails with the theoretical views on the origins of the LFP oscillations as on a result of 
synchronization of the neuronal spiking activity in both the excitatory and inhibitory networks \cite{Strogatz,Arenas,
Hoppensteadt}. Broadly speaking, it is believed that the LFP rhythms are due to a coupling between the electromagnetic 
fields produced by local neuronal groups \cite{Buzsaki1}. If the coupling between these groups is sufficiently high, then 
the individual fields oscillating with amplitudes $a_p$ and phases $x_p$ synchronize, yielding a nonzero mean field 
$\Sigma_p a_p e^{ix_p} = Ae^{i\phi}$ that is macroscopically observed as LFP \cite{Strogatz,Arenas,Hoppensteadt}. 
In particular, the celebrated Kuramoto Model \cite{Strogatz} describes the synchronization between oscillators via a 
system of equations
\begin{equation}
\partial _t x_q = \omega_{q,0} + K\Sigma_p \sin(x_q - x_p),
\label{Kuram}
\end{equation}
according to which the oscillators transit to a synchronized state, as the coupling strength $K$ increases. The Eqs. 
(\ref{Kuram}) directly point out that the synchronized frequency, $\omega(t) = \partial_t \phi$, should have the form 
(\ref{omt}). However, this form of expansion has not been previously extracted from the experimental data, which may 
be due to the fact that the Fourier method does not resolve the spectral structure in sufficient detail (Fig.~\ref{SFigure2}). 
In contrast, the description of the LFP oscillations produced by the DPT method may provide such resolution and help to 
link the empirical data to theoretical models of neuronal synchronization. 

\section{Mathematical Supplement}
\label{section:Supplement}

\begin{figure}[ht] 
\includegraphics[scale=0.72]{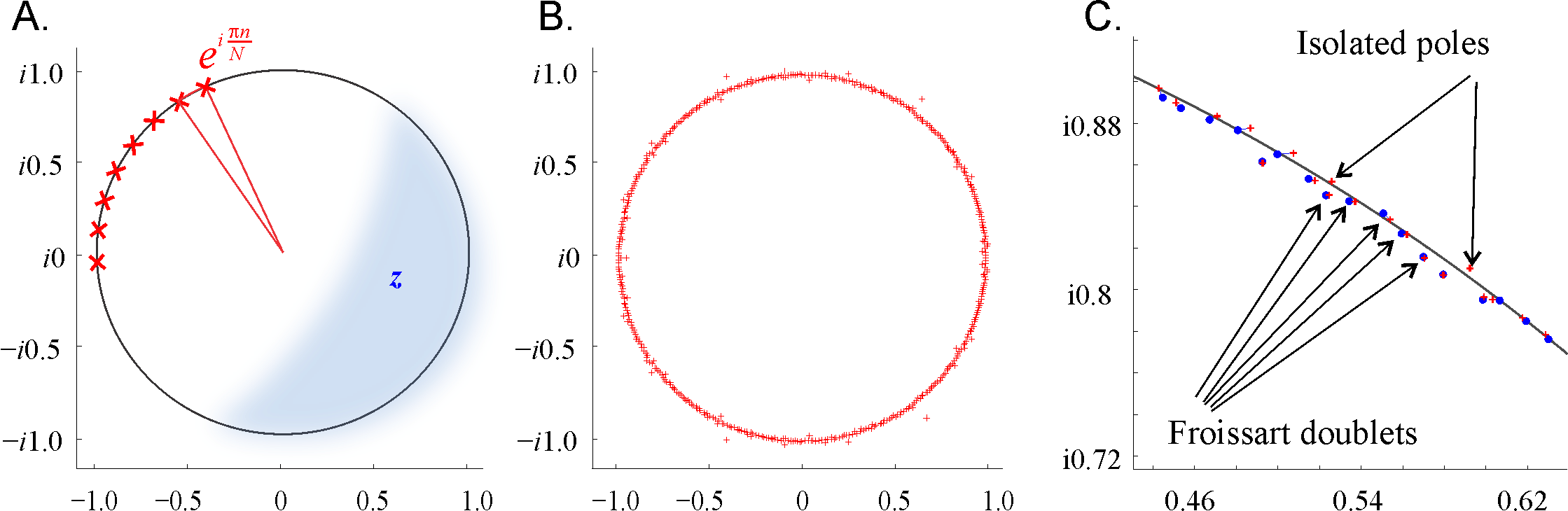} \caption{{\footnotesize \textbf{Fourier and Pad\'{e} in complex plane}. 
\textbf{A}. The discrete waves, $z_l  = e^{i2\pi l/N}$, used to construct the Fourier decompositions, are uniformly 
distributed over the unit circle $S^1$, embedded into the complex plane of the variable $z$. \textbf{B}. The poles 
of the Pad\'{e} approximant to the signal's generating function, $z_p$ (red crosses), also concentrate in a close 
vicinity of the unit circle, in accordance with Steinhaus' theorem \cite{Steinhaus}. For illustrational purposes, the 
number of Pad\'{e}-poles shown on panel B is much larger than the number of harmonics shown on panel A. 
These poles are not constrained to any \emph{a priori} selected locations; in fact, their positions in the complex 
plane $C^{1}$ are dictated solely by the signal's structure, which ultimately leads to the super-resolution property 
\cite{Perotti1,Perotti2,Perotti3}. \textbf{C}. According to the Froissart's theory \cite{Froissart,Bessis2}, zeros and 
poles that represent the noise component of the signal form close pairs---the Froissart doublets. A zoom-in into a 
small segment of the unit circle shows many Froissart doublets (zeros shown as blue dots), and two isolated poles 
that represent the regular, oscillatory part of the signal.}}
\label{SFigure1} 
\end{figure}

\textbf{The Discrete Fourier Transform (DFT)} is used to represent a given time series as a superposition of discrete 
harmonics with \emph{fixed} frequencies (Fig.~\ref{SFigure1}A). To that end, $N$ recorded values, $s_1, s_2, \ldots, 
s_{N}$, are convolved with a set of $N$ discrete harmonics, $z_l = e^{i2\pi l/N}$,
\begin{equation}
A_l = \Sigma_n s_n z_{l}^{n}.
\label{ft}
\end{equation}
The magnitude of this convolution defines the amplitude of the discrete plane wave $z_l$ in the discrete Fourier 
decomposition: the most prominent oscillatory components produce peaks in the Fourier transform, whereas the 
noise broadens these peaks, lowers their magnitudes and generally obscures the spectral properties of the signal 
\cite{Grunbaum}. Similar effects are produced by the signal's nonstationarity, i.e., by the time dependence of the 
signal's frequencies.

\textbf{The Discrete Pad\'{e} Transform (DPT)} method discussed here is based on studying the so-called 
$z$-transform of the recorded time series,
\begin{equation}
S(z) = \Sigma_n s_n z^n,
\label{z}
\end{equation}
where $z = x + iy$ is a complex variable (i.e., the series expansion (\ref{z}) is an extension of (\ref{ft}) into the 
entire complex plane), and of its Pad\'{e} approximant---a ratio of two polynomials $P_{N-1}(z)$ and $Q_N(z)$, 
\begin{equation}
S_N(z) = P_{N-1}(z)/Q_N(z)
\label{pda}
\end{equation}
that approximates $S(z)$ to the $2N$-th order of $z$ \cite{Baker}. 

\emph{Oscillatory component}. 
In the analyses of the oscillatory signals, the $N$ roots $z_p$, $p=1,...,N$, of the polynomial $Q_N(z)$---the poles 
of the Pad\'{e} approximant---play the role of the discrete Fourier harmonics, $z_l$, in the DFT: they capture the 
spectral structure of the signal \cite{Bessis1,Bessis2,Perotti1}. Indeed, consider a signal $r(t)$ obtained as a 
superposition of $N_P$ damped oscillators,
\begin{equation}
r(t) = \Sigma_p A_p e^{-\alpha_p t}\cos(\omega_p t + \varphi_p),
\label{rt}
\end{equation}
where $A_p$ is the amplitude of the $p$th oscillator with a damping exponent $\alpha_p$, frequency $\omega_p$ 
and phase $\varphi_p$. If the signal is sampled at a frequency $S$, then it generates a discrete time series,
\begin{equation}
r_k = \Sigma_{p=1}^{N_{P}} c_p e^{i\omega^{(+)}_p k/S} + c^{*}_p e^{i\omega^{(-)}_p k/S}, \ \ \ k \in \mathbb{Z},
\label{rk}
\end{equation}
where $\omega^{(\pm)} = i \alpha_p \pm \omega_p$ and $ c_p = A_p e^{i\varphi_p /2}$. The generating function 
of this series,
\begin{equation}
R(z) = \Sigma_{k=1}^{\infty} r_k z^k =  \Sigma_{p=1}^{N_{P}} \left(\frac{c_p}{1-z e^{i\omega^{(+)}_p /S }} + 
\frac{c^{*}_p}{1-z e^{i\omega^{(-)}_p /S}}\right),
\label{R}
\end{equation}
is a rational fraction of degree $(2N_P-1)/2N_P$ with poles $$z_{p}^{(\pm)}= e^{-i\omega_p^{(\pm)}/S}.$$
The phase of each pole $z_p$ defines the frequency $\omega_p$, and its magnitude defines the damping constant 
$\alpha_p$. The residue of $z_p$ defines the amplitude $A_p$ and the phase $\varphi_p$ of the corresponding 
oscillator. Notice that poles come in complex conjugate pairs and have to lie either outside of the unit circle, if the 
signal is damped ($\Im \omega_p = \alpha_p > 0$), or on the circle if the signal has no damping.

\emph{Noise component}. If a signal is perturbed by an additive noise $\xi(t)$, then the generating function (\ref{pda}) 
of the resulting ``noisy'' time series $s_n = r_n + \xi_n$ is the sum of the ``regular'' and the ``noisy'' part, $S(z) =
 R(z) + \Xi(z)$, where 
\begin{equation}
\Xi(z) = \Sigma_n \xi_nz^n.
\label{Xi}
\end{equation}
A remarkable theorem proven by H. Steinhaus \cite{Steinhaus} establishes that the poles of $\Xi(z)$ concentrate, 
with probability $1$, at the unit circle (Fig.~\ref{SFigure1}B). In other words, the generating function of a random 
data series is an analytic function inside the unit disk, possessing a dense set of poles as $|z|$ approaches 1. Thus, 
the total generating function of the full signal $S(z) = R(z)+ \Xi(z)$ has a finite number of poles contributed by $R(z)$ 
and an infinite number of poles contributed by $\Xi(z)$. 

%%%%%%%%%%%%%%%%%%%%%%%%%%%%%%%%%%%
%%%%%%%%%%%%%%%%%%%%%%%%%%%%%%%%%%%%

A key property of the Pad\'{e} approximant to $\Xi(z)$ is that its poles occur in close vicinities of its zeros, forming 
the so-called \emph{Froissart doublets} \cite{Froissart,Gilewicz1,Gilewicz2} that can be easily detected numerically 
(Fig.~\ref{SFigure1}C). In our analyses, the typical distance in a pole-zero pair is smaller than $10^{-6}-10^{-7}$ in 
the standard Euclidean metric on $C^1$. We hence identified such pairs as the ones smaller than a critical distance 
$\delta =10^{-5}$. These results are stable: injecting small amounts of white and colored noise into the signal (about 
$10^{-3}$ of the signal's mean amplitude, i.e., at least ten times more than the signal's natural noise level) does not 
alter the reconstructed positions of the regular poles and hence the parameters the spectral waves remain the same 
as the ``perturbed'' Froissart doublets are removed. 

\textbf{$J$-matrix formalism}. In order to obtain a Pad\'{e} approximation to $S(z)$ in the entire complex plane, the 
$\Xi(z)$ has to be analytically extend through its natural boundary, which remains an open problem of complex analysis. 
However, the ``$J$-matrix approach'' developed in a recent series of publications \cite{Bessis1,Bessis2,Perotti1}, allows 
addressing this problem in practical terms. The generating function $G(z)$ can be associated with a tri-diagonal Hilbert 
space operator $J$ that has $G(z)$ as its resolvent matrix element, $G(z) = \langle e_0|(J-z1)^{-1}|e_0\rangle$, $e_0= 
(1,0,...)$ \cite{Baker}. In accordance with Steinhaus' theorem, the spectrum of $J$ consists of two parts: an essential 
spectrum with support on the unit circle, which represents the noise component and a discrete spectrum, containing a 
finite number of poles outside the unit circle, which represent the regular component of the signal (a finite number of 
damped oscillators). In the spectrum of finite order truncations $J_N$ of the $J$-operator, the poles of the Froissart 
doublets take the place of the essential spectrum. Moreover, these finite matrices can be explicitly constructed as follows. 
Let us consider the set of subdiagonal Pad\'{e} approximations to the generating function of a given time series defined 
by (\ref{pda}). The polynomials $Q_N(z)$ satisfy a third order recursive relation which can be written in a matrix form, 
$J_N V = zV$ where $J_N$ \emph{is the (tri-diagonal) finite order matrix approximation to the $J$-operator of order $N+1$}. 
The column vector $V$ is defined by the polynomials $Q_N$, $V_T = [Q_0(z), Q_1(z), ..., Q_N(z)]$. The zeros of $Q_{N+1}(z)$ 
define the eigenvalues $(z_0, z_1, ..., z_N)$ of $J_N$ and therefore the poles of $R_N$. The same procedure applied to 
$P_N$ (with a slightly modified matrix) gives us the zeros of 
$G_N$, thus completely characterizing $R_N$ itself.

\textbf{Short Time Pad\'{e} Transform, (STPT)} is analogous to the standard Short Time Fourier Transform (STFT) 
method \cite{Jacobsen}. Starting with a segment $s_1, s_2, ..., s_N$ centered at $t_1$, we compute the Pad\'{e} 
approximants, identify and discard the Froissart doublets, and then evaluate the frequencies, $\omega_q(t_1)$, the 
amplitudes, $A_q(t_1)$ and the phases, $\varphi_q(t_1)$, associated with the stable poles $z_1, z_2, ...z_{p_1}$, 
$q = 1, ..., p_1$. After that, the window is shifted by $\Delta T$ to the position centered at $t_2$, and the same analysis 
is applied to the next segment of the time series, revealing the frequencies, $\omega_i(t_2)$, the amplitudes, $A_i(t_2)$ 
and the phases, $\varphi_i(t_2)$, $i = 1, ..., p_2$, and so on. 

%By construction, this difference is produced by the discarded Froissart doublets, i.e., by the noise component of the signal. 
%Thus, the combined contribution of the ``noisy'' poles is very small even though their number is large, which implies that 
%the brain waves, according to the DPT method, are virtually noiseless. In other words, the DPT method suggests that the 
%standard combination of Fourier plane waves with \emph{constant} frequencies,
%\begin{equation}
%r(t) = \Sigma^N_{p =1} A_p e^{i\omega_pt},
%\label{fourdec}
%\end{equation}
%can be replaced by a combination of a few frequency-modulated waves,
%\begin{equation}
%r(t) = \Sigma^M_{q=1} A_q e^{i\phi_q(t)},
%\label{oscdec}
%\end{equation}
%$N \gg M$, which we call \emph{oscillons}. 

\textbf{Separating the noise from the oscillations}. The Froissart doublets and the regular poles of $S(z)$, exhibit 
qualitatively different behaviors in response to changes of the DPT algorithms' parameters. If the size of time window 
$T_W$ in the STPT is altered, or as it is shifted from one segment of the time series to another, or if the order of the 
Pad\'{e} approximant is changed, the Froissart-paired poles move significantly and irregularly around the unit circle, 
as one would expect from a structure that represents noise. In contrast, the poles associated with the regular part of 
the signal remain stable and isolated. These differences can be easily detected numerically, producing the computational 
DPT method \cite{Bessis1,Bessis2}. 

Note that every data point obtained by the STPT method is obtained independently: evaluation of $N$ frequencies at 
each time-step, identification of the ``noisy'' vs. ``regular'' frequencies, etc. does not affect the values obtained at the 
other time-steps and hence the pattern formed by the data points is completely empirical. 

\textbf{Computing the parameters of the spectral waves}. Since the instantaneous parameters of the oscillons are 
computed independently based on a finite number of data points, the reconstructed spectral waves contain gaps and 
other irregularities. We therefore construct the smoothened spectral waves by interpolating the ``raw'' traces of the
regular frequencies over the uniformly spaced time points, and compute the mean parameters $\omega_{q,0}$, $\omega_{q,i}$, 
$\Omega_{q,i}$, and $\varphi_{q,i}$ in the expansion
\begin{equation}
\omega_q (t) \equiv \partial_t \phi_q = \omega_{q,0} + \omega_{q,1} \sin(\Omega_{q,1}t + \varphi_{q,1}) + 
\omega_{q,2}\sin(\Omega_{q,2} t +\varphi_{q,2}) + \ldots.
\label{omgq}
\end{equation}
using the standard DFT methods. 

\begin{figure}
	\includegraphics[scale=0.72]{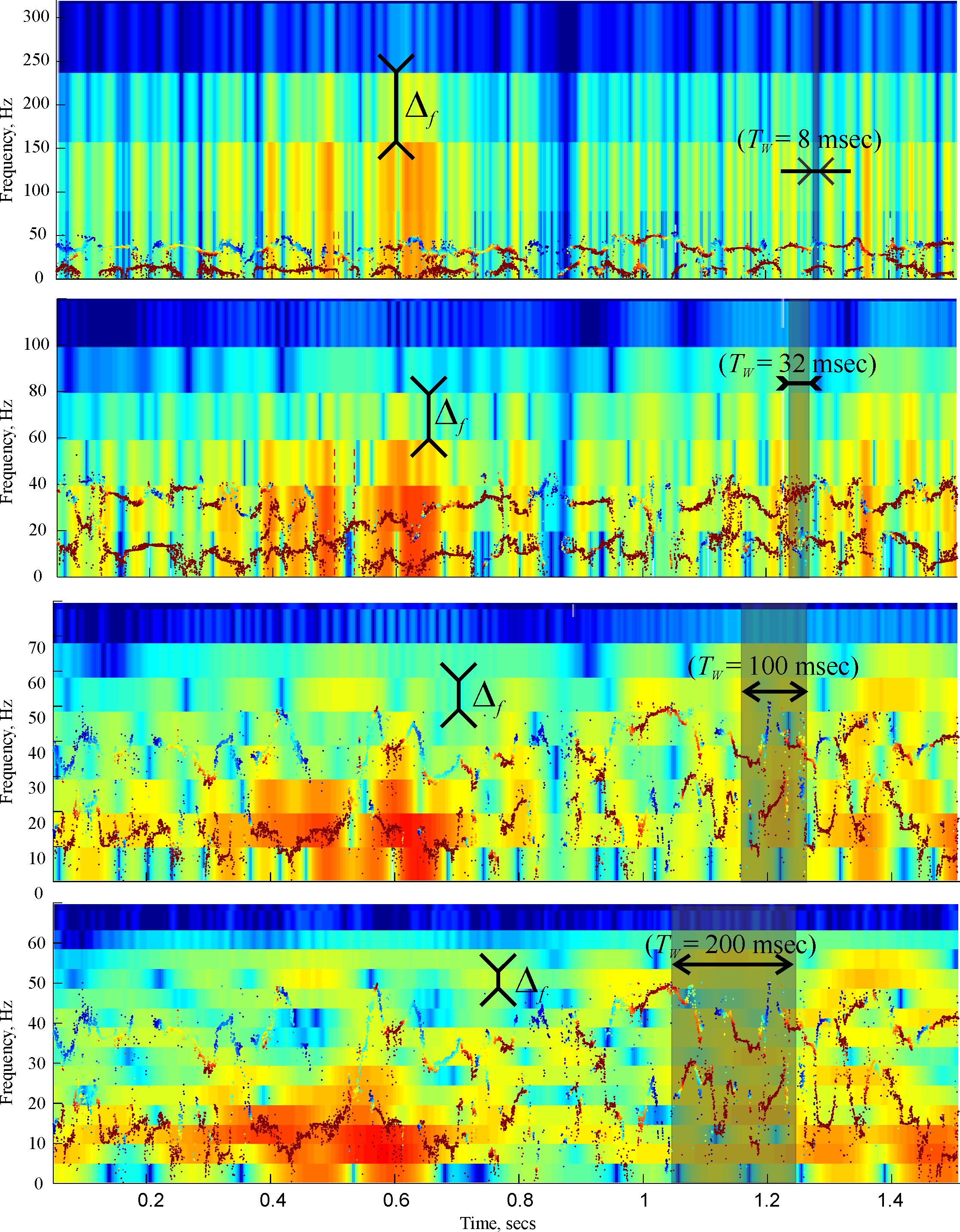}
	\caption{{\footnotesize \textbf{Insufficiency of Fourier resolution}. The two lowest ($\theta$ and low-$\gamma$) spectral 
			waves of an LFP signal, filtered between $1$ and $56$ Hz, are superimposed on four Fourier spectrograms of the 
			same signal, computed for $T_W=8$ msec, $T_W=32$ msec, $T_W=100$ msec and $T_W=200$ msec. The sliding 
			window widths $T_W$ are shown by gray vertical stripes. The horizontal stripes on each Fourier spectrogram indicate 
			the magnitude of the spectral resolution, $\Delta_f$. The spectral resolution of the Fourier spectrogram becomes 
			comparable to the frequency scale of the spectral waves only for $T_W = 100$ msec (third panel from the top), but 
			the temporal resolution at this value exceeds the characteristic period of the spectral waves. Increasing the frequency 
			resolution broadens the window size beyond the spectral waves' period (bottom panel) and vice versa, increasing 
			temporal resolution destroys the frequency resolution (top two panels). As a result, the spectral waves remain 
			unresolved by the DFT, which can only detect a band of increased amplitudes, but not the detailed pattern of the 
			oscillating frequencies.}}
	\label{SFigure2}
\end{figure}

The Fig.~\ref{SFigure2} illustrates how the inherent conflict between the time and the frequency resolutions 
obscures the spectral waves in  the Fourier spectrogram. 
\clearpage

%%%%%%%%%%%%%%%%%%%%%%%%%%%%%%%%%%%
%%%%%%%%%%%%%%%%%%%%%%%%%%%%%%%%%%%%

\begin{figure}[h!]
	\includegraphics[scale=0.72]{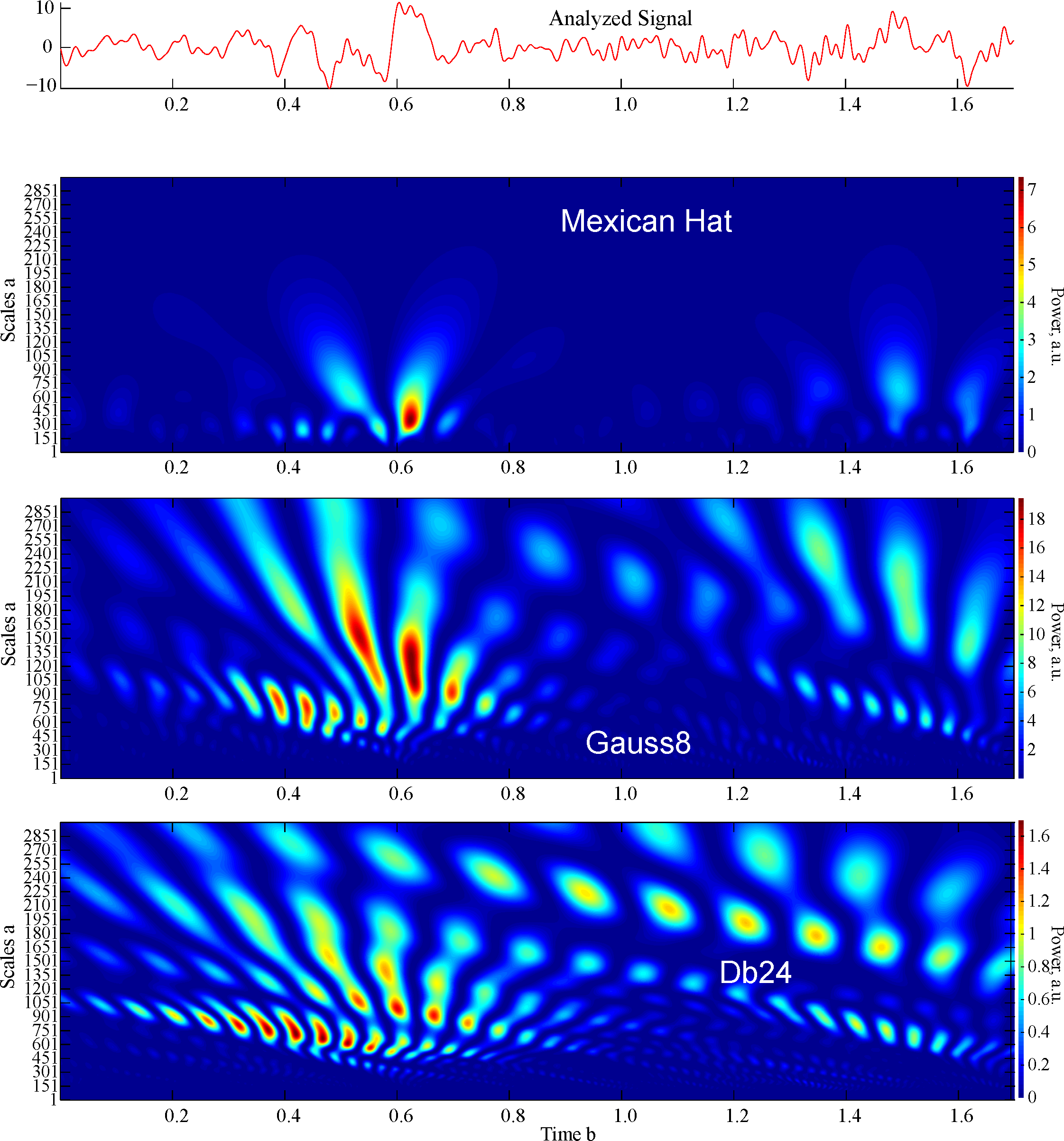}
	\caption{{\footnotesize \textbf{Wavelet spectrograms} of the same signal, computed for the ``Mexican hat'' (top panel), the 
	Gauss wavelet of level 8 (middle panel) and Daubechies' wavelet of level 24 (bottom panel). In all three cases,the maxima of 
	the wavelet coefficients correspond to undulating patterns of the signals' amplitude at different temporal scales but do not 
	resolve the spectral waves.}}
	\label{SFigure3}
\end{figure}

\clearpage
The wavelet spectrograms (scalograms) of the same signal using three different wavelets are shown on Fig.~\ref{SFigure3}. 
 
%%%%%%%%%%%%%%%%%%%%%%%%%%%%%%%%%%%
%%%%%%%%%%%%%%%%%%%%%%%%%%%%%%%%%%%%

%%%%%%%%%%%%%%%%%%%%%%%%%%%%%%%%%%%
\begin{figure} 
	\includegraphics[scale=0.72]{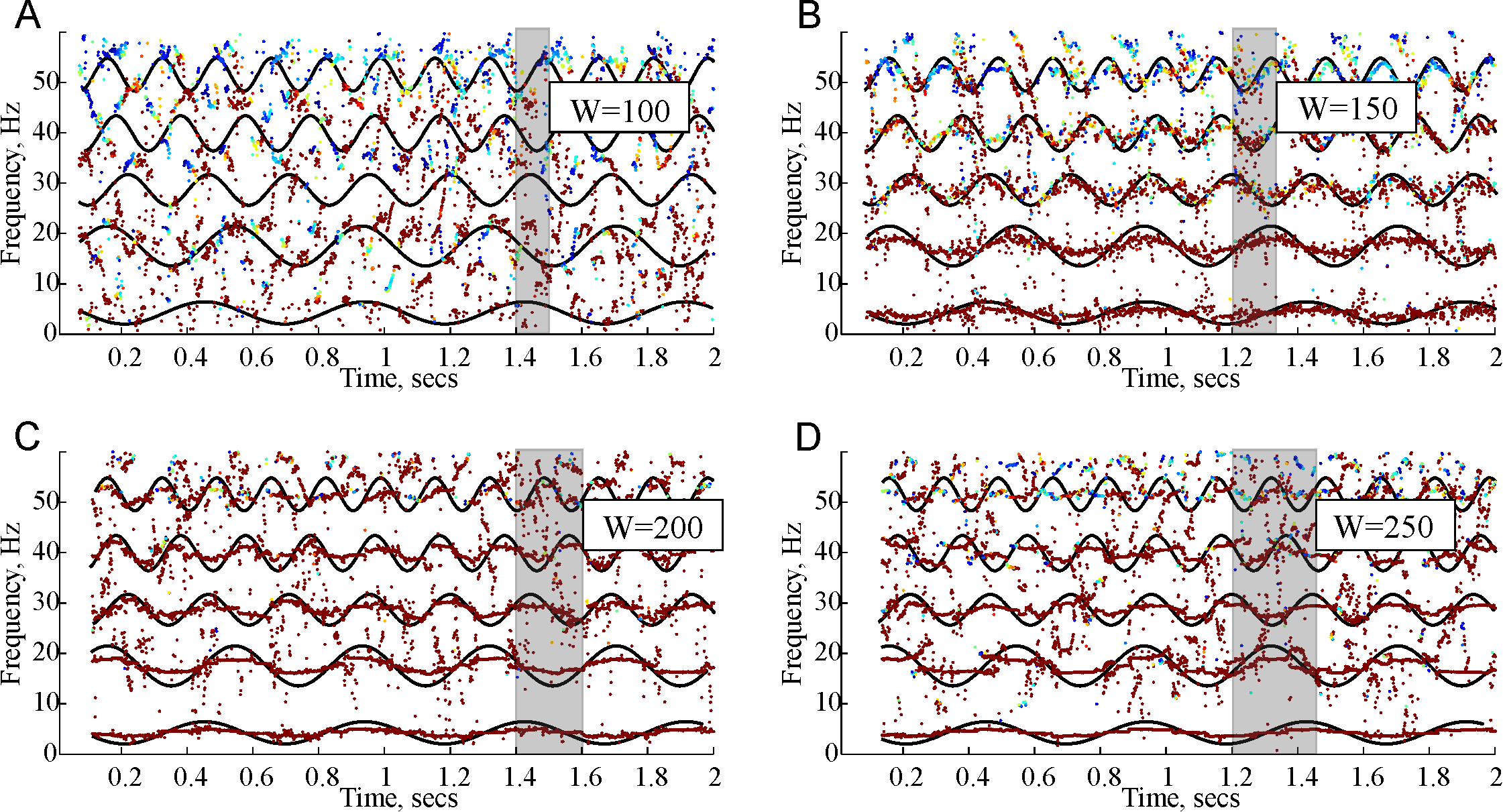}
	\caption{{\footnotesize \textbf{Discrete Pad\'{e} spectrograms} of the simulated combination of five oscillons, computed for 
			four window widths ($T_W = 0.10$ sec, $T_W = 0.15$ sec, $T_W = 0.2$ sec and $T_W = 0.25$) . While the first 
			window is too narrow to capture the structure of the spectral waves, the second window resolves them. The last 
			two values of $T_W$ are too large--the undulating pattern of the spectral waves is replaced by the emerging 
			sidebands.}}
	\label{SFigure4}
\end{figure} 
%%%%%%%%%%%%%%%%%%%%%%%%%%%%%%%%%%%%

To illustrate the effectiveness of the proposed method, we simulated a superposition of five artificial oscillons with the amplitudes 
$A_1 = 0.5$, $A_2 = 0.3$, $A_3 = 0.15$, $A_4 = 0.1$ and $A_5 = 0.05$, the mean frequencies are $\omega_{1,0}=5$ Hz, 
$\omega_{2,0}\approx 20$ Hz, $\omega_{3,0}\approx30$ Hz, $\omega_{4,0} \approx 40$ Hz, and $\omega_{5,0} \approx 50$ 
Hz, and five modulating frequencies $\omega_{1,1}=2$ Hz, $\omega_{2,1}=2.5$ Hz, $\omega_{3,1}=4$ Hz, $\omega_{4,1}=5$ 
Hz and $\omega_{2,1}=6$ Hz respectively. The amplitudes of the frequency modulations are approximately $\pi$ Hz in all cases. 
The resulting ``spectral waves'' are shown as black sinusoids in the background of the four panels of Fig.~\ref{SFigure4}. 
Each panel corresponds to a particular window width: $T_W = 0.10$ sec, $T_W = 0.15$ sec, $T_W = 0.2$ sec and $T_W = 0.25$ sec,
at the sampling rate of $S=1000$ Hz. In the first case, the DPT is therefore based on $N_1 = 100$ data points per window, i.e., $50$ 
sample frequencies occupying the range between $0$ and $500$ Hz, or about one frequency per $10$ Hz interval. As shown on the 
Fig.~\ref{SFigure4}A, this (or smaller) values are insufficient for resolving oscillons with magnitude $\pm\pi$: the undulatory pattern 
is not captured. 
In the second case, each window contains $N_2 =150$ points, or one frequency per $\approx 6.6$ Hz, and the spectral waves become 
apparent (Fig.~\ref{SFigure4}B). If window becomes bigger, $N_3 =200$ points (Fig.~\ref{SFigure4}C) or $N_4 = 250$ points 
(Fig.~\ref{SFigure4}D), the temporal resolution suffers: the patterns of all spectral waves become averaged over the window width. 
As a result the upper spectral waves produce sidebands and the lowest spectral waves the flatten out. 
We emphasize however, that the original signal can be reconstructed with high precision in all cases; the issue is only whether the 
spectral can or cannot be resolved.

\begin{figure} 
	\includegraphics[scale=0.72]{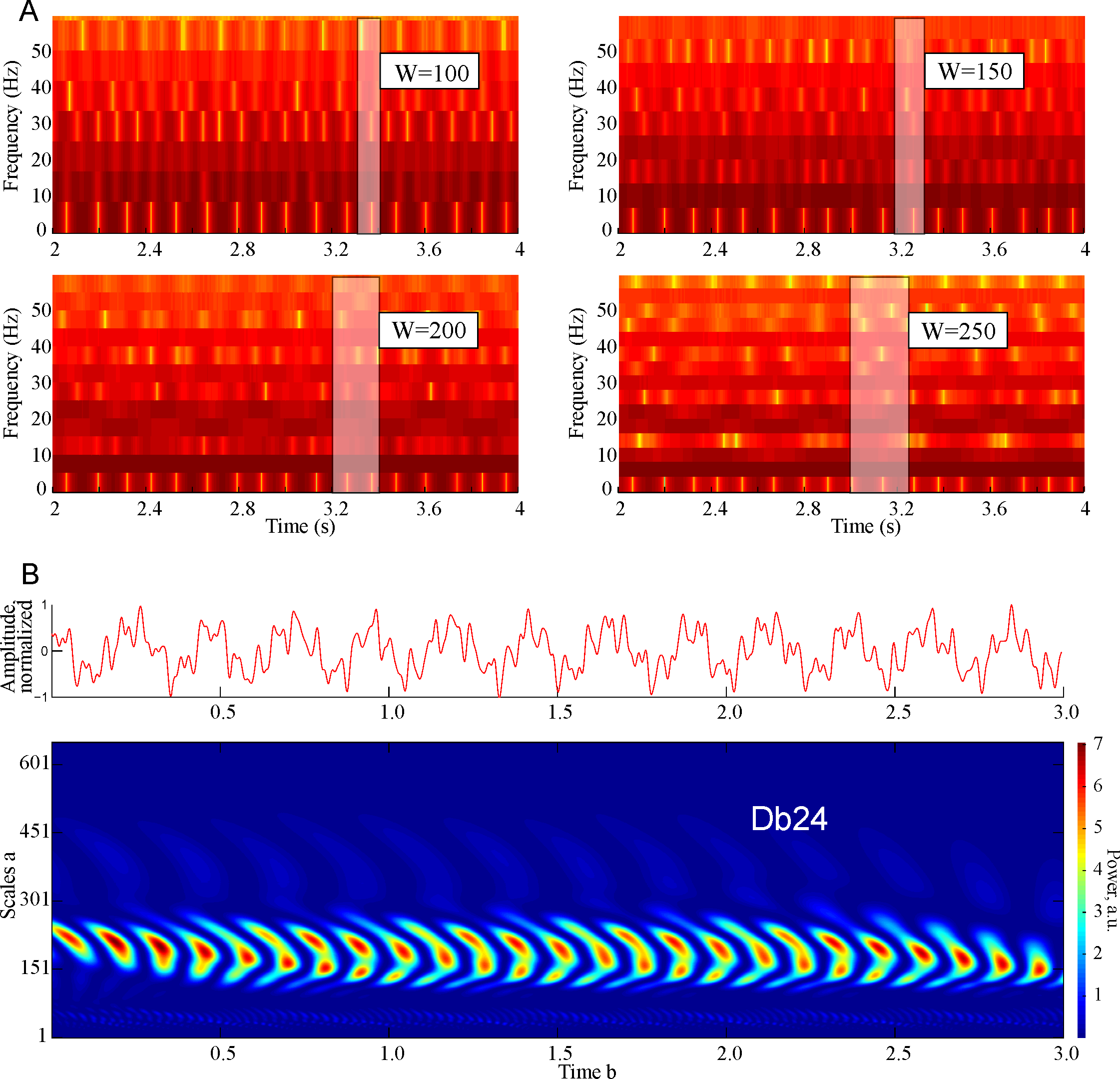}
	\caption{{\footnotesize \textbf{Alternative methods}. \textbf{A}. The Fourier spectrograms of the same signal as shown on
	 Fig.~\ref{SFigure4}, computed for the same window widths ($T_W = 0.10$ sec, $T_W = 0.15$ sec, $T_W = 0.2$ sec and 
	 $T_W = 0.25$) do not resolve the simulated spectral waves. \textbf{B}. Wavelet spectrograms of the same signal, computed 
	 using Daubechies' wavelet of level 24, also captures the undulatory pattern of the signal but does not resolve the spectral waves.}}
	\label{SFigure5}
\end{figure}

For comparison, the corresponding Fourier spectrograms and the wavelet scalogram computed using Daubechies' wavelet of level 24 
are shown on Fig.~\ref{SFigure5}.

%%%%%%%%%%%%%%%%%%%%%%%%%%%%%%%%%%%%
%%%%%%%%%%%%%%%%%%%%%%%%%%%%%%%%%%%%

\clearpage
%\newpage

\section*{Acknowledgments}
\label{section:acknow}

The authors thank Dr. A. Tsvetkov for valuable discussions, Dr. J. Tang from the Baylor College of Medicine for providing 
LFP signals for this study. The work was supported by the NSF 1422438 grant (Y.D.), by NSF grants 0114796 (2001-2003), 
DBI-0318415 and DBI-0547695 (2006-2009) (L.P. and D.B.).

\end{document}